\documentclass[11pt,twoside]{article}
\usepackage{amsmath,amssymb}
\usepackage[all]{xy}
\usepackage{color}
\usepackage[hmargin=2cm,vmargin=2.5cm]{geometry}

\begin{document}

\title{Spectrally determined singularities in a potential with an inverse square initial
term}
\author{Demetrios A. Pliakis\\
Department of Electronics,TEI Crete\\ Romanou 3, 
Chalepa, Chania, GR-73133\\
{\ttfamily pliakis@chania.teicrete.gr}}
\maketitle

\begin{abstract}
We study the inverse spectral problem for Bessel type operators with
potential \(v(x)\): \( H_\kappa=-\partial_x^2+\frac{k}{x^2}+v(x)\). The
potential is assumed smooth in \((0,R)\) and with an asymptotic expansion in
powers and logarithms as
\(x\rightarrow 0^+,  v(x)=O(x^\alpha), \alpha >-2\).
Specifically we show that the coefficients of the asymptotic expansion
of the  potential are spectrally determined. This is achieved by computing
the expansion of the trace of the resolvent of this operator which is
spectrally determined and elaborating the relation of
the expansion of the resolvent with that of the potential, through the
singular asymptotics lemma.

\end{abstract}

\section{Introduction}

\vspace{0.5truecm}
\noindent Let $R>0, \alpha >-2$ and $v$ be a real valued function, smooth 
in $C^\infty_0((0,R))$ that has an asymptotic expansion together with all
its derivatives as $x \rightarrow 0^+$ of the form 
$$v(x) \sim x^\alpha \sum_{n=0}^\infty v_n x^{k_n} $$ where
$\{ k_n \}$ is an increasing sequence of positive real numbers. 
This function  appears in the singular Sturm-Liouville operator 
$H=-\partial_x^2+\frac{\kappa}{x^2}+v(x)$ 
on $(0,R)$ with Dirichlet boundary conditions at $0,R$. 
In this paper we examine  whether the spectrum of this Sturm-Liouville 
operator uniquely determines the sequence of asymptotic coefficients 
$\{ v_n \}$ of the potential function.

\vspace{0.5truecm}
\noindent  We prove that this is indeed the case 
provided that $v_0 \neq 0$; this is achieved by
the explicit construction of the  spectral invariants that in turn 
determine the asymptotic coefficients of the potential function $v$ 
from the spectrum. 
The function $$f_H(\eta)=\eta^2Tr(1+\eta^2H)^{-1}$$ is defined for small
$\eta >0$ and is of course spectrally determined. The $\eta \rightarrow 0$
asymptotic expansion of $f$  will provide the desired spectral 
invariants: the singularity of the Sturm -Liouville operator $H$ manifests
itself in the appearance of coefficients in the $\eta \rightarrow 0^+$ asymptotic expansion  
of $f_H(\eta)$ that depend polynomially on the coefficients $\{ v_n\}$ of
$v$. Precisely we have that this  asymptotic expansion is of the form 
\begin{equation}
f_{H}(\eta)\sim \sum_{n=0}^{\infty}A_n\eta^{2n+1} + 
\eta^4(\sum_{n=-1}^{\infty}B_{n}\eta^{2n+1}\log\eta+
\sum_{n=0}^{\infty}C_n\eta^{l_n})
\end{equation} where $l_n$ is real but not an odd integer. Evidently it contains apart from pure odd powers also new terms, odd powers with logarithms  as well as pure powers other than the odd ones. 
The sequences of asymptotic coefficients $B_n,C_n$ of the 
asymptotic expansion of $f$ are polynomials in the asymptotic coefficients
of the potential $\{ v_n \}$; it is exactly this fact that allows us to determine recursively these 
$\{ v_n \}$ from the ''heat invariants'' that are equivalent to the $\eta
\rightarrow 0^+$ asymptotic coefficients of $f_H(\eta)$. The exact
formulation of this tehorem is the following:

\vspace{0.5truecm}
\noindent
{\bf Theorem.} {\it Let $\chi$ be a smooth cutoff function, $\chi \equiv 1$
in a neighborhood of the origin, then the function $f_{\chi,H}(\eta)$
has an asymptotic expansion of the form 
$$ f_{\chi,H}(\eta) \sim \sum_{k=0}^{\infty}A_{2k+1}\eta^{2k+1}+
\sum_{n=0}^{\infty}
B_{n}\eta^{z_n}+   
\sum_{k=0}^{\infty}  C_{2k}\eta^{2k} +
 \sum_{k=0}^{\infty}  D_{2k+1}\eta^{2k+1}\log \eta $$
where $z_n \in {\bf R}\setminus {\bf Z},B_{k},C_{k},D_{k}$ are homogeneous
polynomials in the asympotic coefficients $\{v_n\}$ of the potential 
of degree $z$ if we define $\mbox{deg}(v_n)=k_n+2$.} 

\vspace{0.5truecm}
\noindent

Inverse spectral results of the form we described above appeared first in
\cite{c1},\cite{c2},\cite{c3}, \cite{cu},\cite{ct}. Actually in \cite{ct}
certain coefficients are calculated in the course of calculations of
operator determinats and in the rest the generic case is studied
exhaustively. The general case requires the treatment of certain exceptional cases.
This treatment is provided here and  leads to  the full result.

The conical singularities have been treated extensively in the literature for 
diverse purposes giving rise to diverse results and calculations. We refer
to the first papers that dealt with asymptotics of either the heat 
\cite{ch2} or the wave kernel \cite{cht} on spaces with such singularities. 

Possible applications of these results are indicated already in 
\cite{c3}. We recall  these briefly here. Potentials of this type arise 
in the wave equation for a vibrating rod of variable cross section, when 
the cross-sectional area of the rod vanishes quartically
(as a function of the distance from the end of the rod) at one
point. In the same spirit we could determine the profile of a surface of 
revolution with a conical singularity on the axis, asymptotically to all 
orders, from the spectrum of Laplacian restricted to functions with polar 
symmetry. 

We indicate here briefly two applications in physics. 
Furthermore the present study exhausts the determination of  
a confining potential
which is hydrogen atom-like at short distance \cite{gm}, from the complete set
of bound state energies. This is indicated in \cite{c1} but only for the case 
of the $s-$wave. Hence the present study fills in the higher angular momemtum 
instances  for potentials of asymptotic expansion of  the form given above.
We will present here briefly the construction in \cite{c1}. Confinement is
expressed by Dirichlet boundary conditions on the surface of a sphere. The
Hamiltonian with a radial potential is
$$ H= -\Delta_3+v(|x|) $$ where $v \in \Gamma^{\infty}((0,R))$ with
$v(x)=O(x^{\alpha}), \alpha >-2$ as $x \rightarrow 0$. The boundary 
condition on the domain of $H$ is $\psi(R)=0$. The $l-$th  spherical 
harmonic Hamiltonian is unitarily equivalent  to
$$H_0=-\partial_r^2+\frac{l(l+1)}{r^2}+v(r)$$ which is an an unbounded 
operator on $L^{2}((0,R);dr)$. The boundary conditions inherited from the 
3-dimensional problem are $f(0)=f(R)=0$. A physically meaningful question
is whether the potential $v$ could be determined from the mass spectrum of the 
bound states associated to a given spherical harmonic because both the 
mass (the energy) and the angular momentum (the order of the spherical 
harmonic) are measurable in the laboratory.

The study we perform answers this
affirmatively when the potential is a real analytic function on $(0,R)$ and
that near ${0}$ is written as a convergent series of the following form
$$v(x)=x^{-2}\sum_{k,j}v_{k,j}x^{\alpha_k}\log^j x,$$  for
$\{ \alpha_k \}$ an increasing sequence of positive real numbers.
If the function $v$ is not of this form then we can only obtain
the full asymptotic expansion of the potential at small distances.
Moreover we can deduce several corollaries of the preceding theorem that
are interesting in physics. For instance availing ourselves  of the results
of \cite{c5} on irregular singularities, we state the following:

\smallskip

{\bf Corollary.} {\it Let $H=-\partial_x^2+v(x)+P(x)$ be an oprator on ${\bf R}_+$
with Dirichlet boundary conditions. Let $v$ be  a real valued analytic function,
rapidly decreasing at $\infty$, while near $0$ it is given by
a convergent series of the form
$$v(x)=x^{-2}\sum_{k,j}v_{k,j}x^{\alpha_k}\log^j x $$  for
$\{ \alpha_k \}$ an increasing sequance of positive real numbers. The
function $P$ is given and is of the form $$P(x)=\sum_{n=0}^Na_kx^{\beta_n}$$
for $0\leq \beta_0 < \dots < \beta_N$ and $a_N>0$. Then the spectrum of
$H$ determines $v$. }

\smallskip

The preceding result incorporates for instance  the
case of quantum particle systems trapped in magnetic fields. An immediate \
application of the preceding corollary refers to the
2-dimensional magnetic trap \cite{ahs1}, due to a vanishing magnetic field
at $\infty$. Precisely,  let $0<\epsilon<1$ and $(r,\theta)$ be the
coordinates in ${\bf R}^2$ and consider the quanutm motion in
 a magnetic field of the type $B=(2-\epsilon)r^{-\epsilon}$. The
 Hamilotnian for a particle  subject
to such a magnetic field as well as to an unknown radial potential $v(r)$
has the form $$H=-\partial^2_r+\frac{1}{r^2}\partial_\theta^2+
r^{2(1-\epsilon)}+2ir^{-\epsilon}\partial_\theta +v(r)$$ and restricting to
angular momentum $m$ we obtain an 1-dimensional operator of the
above type provided that $v(r)$ is of the required form. We notice also the
example from \cite{ll} of a quantum particle moving in a magnetic field
with fixed angular momentum in the direction of the field. Similar
conclusions could be obtained in that case as well.
The results in \cite{ahs2}
allow us to incorporate the case of a pair of opposite charges with
fixed relative angular momentum  moving in a
homogeneous magnetic field and interacting through  an unknown potential that we
 wish to determine spectrally when the latter satisfies the preceding
 assumptions.

 Actually we insert a cutoff function $\chi, supp \chi \subset [0,R) $ in
order to deal the operator in the half line.  The $\eta \rightarrow 0$ asymptotic
expansion of $$ f_{{\tilde H}}(\eta)=\eta^2tr(1+\eta^2{\tilde H})^{-1}$$
- ${\tilde H}$ is the operator in $(0,\infty)$-
coincides $mod(\eta^{\infty})$ terms with that of $f_H$.

The paper is organised as follows: we start with the neccessary
facts concerning the operator domain, the operator estimates  and 
we construct the resolvent of the unperturbed operator. Next we
give the existence and the precise form  of the asymptotic expansion for 
the operator function we introduced above. We conclude with  the 
calculation of the asymptotic coefficients that provide the recursive
relations that determine the potential asymptotic coefficients. The 
solution of these require certain elementary estimates that are included.
 
{\bf Acknowledgements.} I would like to thank my advisor Professor C. Callias
who suggested the problem as well as Professor E. Floratos who pointed out
the possible appliction to the case of magnetic fields. 

\noindent

\section{Operator Domain and Operator estimates}

\subsection{Function spaces for singular heat expansions}
We will consider operators of the form $H=H_{\kappa}+v$, for 
$H_{\kappa}= -\partial_{x}^{2} + \frac{\kappa}{x^2} $ and $v$ is  a real 
valued potential function belongs to the space $ \Gamma^{\infty}(0,R)$ of
functions that have an asymptotic expansion as $x\rightarrow 0^+$ 
together with their derivatives. This space is described  as follows 
that are encountered in the resolvent expansions that we are going to deal
with.

 The space $\Gamma^{\infty}({\bf R}_{+})$
consists of the $C^{\infty}({\bf R}_{+})-$ functions that have asymptotic
expansions as $x \rightarrow 0^{+}$, together with all their derivatives, 
in complex powers of $x$ and integral powers of $\log x$. Precisely, 
$f \in \Gamma^{\infty}({\bf R}_{+})$ if $f \in C^{\infty}({\bf R}_{+})$
and there exists a $S: {\bf C} \rightarrow {\bf Z}_{+}$ such that
$\sum_{\Re z < a} S(z) < \infty $ for each $a \in {\bf R}$, and
$$\partial_x^m f(x)=\sum_{\Re z \leq a}
\sum_{{\bf Z}_{+} \ni j < S(z)}f_{z,j}\partial_x^m(x^z\log^j x) + 
O(x^{a+\delta-m}) $$ for some $\delta=\delta_a >0 $, some $f_{z,j} \in
{\bf C}$ and all $a \in {\bf R}$. $S$ is called the asymptotic character of the 
expansion. Define the following differential operators
on $C^{\infty}(0,\infty),$ for given $S:{\bf C}\rightarrow {\bf Z}_{+}$ as
described above $$P_{z}[S]=\prod_{\Re z' \leq \Re z}(x\partial_x
-z')^{S(z')}, \hspace{0.25truecm}
\overline{P_z}[S]=P_z[S](x\partial_x-z)^{S(z')}= 
\prod_{\Re z' \leq \Re z}(x\partial_x-z')^{S(z')}$$ 
The following proposition is proved in \cite{c4}; though it is elementary 
it provides criterion that  shades light on these spaces 

\vspace{0.5truecm}
\noindent
{\bf Proposition 1.} {\it $f \in \Gamma^{\infty}(0,\infty)$ iff 
$f\in C^{\infty}(0,\infty)$ and there is an asymptotic character such that 
$P_z[S]f(x)=O(x^{z-\varepsilon})$  for all $z\in {\bf C}$ and $\varepsilon
>0$.}

\vspace{0.5truecm}
\noindent
The space $\Gamma^{\infty}(({\bf R}_{+})^2)$ consists of all the functions
$f \in C^{\infty}(({\bf R})^2)$ for which there exist $S_1,S_2$ such that 
$$(x\partial_x)^{s_1}(y\partial_y)^{s_2}\overline{P^x_{z_1}}[S_1]
\overline{P^y_{z_2}}[S_2]f(x,y)=O(x^{\Re z_1+\delta_1}y^{\Re z_2+\delta_2})$$
for all $z_1,z_2 \in {\bf C}, s_1,s_2\in {\bf Z}_{+},(x,y)$ in any
compact neighborhood of $\partial({\bf R}_{+}^2)$ and some $\delta_{j}$ 
depending on $z_1,z_2$.

\noindent
Evidently $(S_1,S_2)$ is analogously the asymptotic character of 
$f \in \Gamma^{\infty}({\bf R}_+^2)$. If $f \in \Gamma^{\infty}({\bf R}_+)$,
we let $D_{k,j}f(0)$ denote the coefficient of $x^k\log^jx$ in the expansion
of $f$ as $x \rightarrow 0$.

Now let  $\alpha >-2, \{k_n \}$ be an increasing sequence of positive real numbers then we have that $S(z)=1$ for $z=k_n+\alpha$ and $S(z)=0$ otherwise.
equivalently, the potential function has the following  asymptotic expansion as $x \rightarrow 0$
 $$v(x) \sim x^{\alpha}\sum_{n=0}^{\infty}v_nx^{k_n}$$

\noindent Actually for $0< x < \epsilon $ and $\alpha >-2$ the obvious estimate 
$ x^{\alpha} \leq \epsilon x^{-2} +B_{\epsilon}$ implies for 
$\phi \in C^{\infty}_0({\bf R}_+)$ and $c <1$ that $$\| V \phi \|_{L^2} \leq 
c \| H_{\kappa} \phi \|_{L^2} + b\| \phi \|_{L^2}$$ which in view of the 
Kato-Rellich theorem reduces the domain questions of $H$ to those for 
$H_{\kappa}$.

\noindent
Through the Hardy inequality we conclude that if $\kappa \geq -\frac14 $ then the operator $H_{\kappa}$ is positive and hence possesses at least one self adjoint extension, the {\it Friedrichs' extension}.  Recall that if $\kappa \geq \frac34 $, 
 then Weyl's criterion implies  that the operator 
$H_{\kappa}$ is essentially self-adjoint and hence the Friedrichs' is the 
only extension. The domain of self adjointness consists of  the functions 
$$ {\cal D}(H)= \{
\phi \in L^{2}(\overline{{\bf R}_+}), \| \partial_{x}^2 \phi \|_{L^2} < \infty,
\| x^{-2} \phi \|_{L^{2}} < \infty \} \subset 
L^2(\overline{{\bf R}_+};\frac{dx}{x^4}) \cap H^2(\overline{{\bf R}_+}).$$ 
The Sobolev embedding theorem implies that these are $L^2$-functions
with absolutely continuous first derivative.  Additionally, since
$\phi \in {\cal D}(H)$ then $\| \frac{\phi}{x^2} \| < \infty $ which in
turn expresses the Dirichlet boundary conditions, 
$\phi(0) = \lim_{x \rightarrow  0} \phi(x)=0$. 

\subsection{The resolvent.}
Let  $H=-\partial_x^2 + v $ be  a Shr\"odinger operator, with real valued 
potential  function. Clearly, it is formally symmetric. If $\phi,\psi$ are the unique 
elements of $kerH$ that are integrable at $0,\infty$ respectively then the 
resolvent of $H$ is given by $$R_{\lambda}(x,y)=
\Theta(x-y)
\frac{\psi(x,\lambda)\phi(y,\lambda)}{W_x(\phi(\lambda),\psi(\lambda))}+ 
\Theta(y-x)
\frac{\phi(x,\lambda)\psi(y,\lambda)}{W_{x}(\phi(\lambda),\psi(\lambda))}$$
where $W$ is the Wronskian of the solutions of 
$(H-\lambda)\phi=0$. In the case of the operator $H_{\kappa,\alpha}= 
-\partial^2_x+\frac{\kappa}{x^2}+\frac{\alpha}{x}$ the solutions are
expressed through  the confluent 
hypergeometric functions \cite{t1}, \cite{gr}:
$$\phi(x,\lambda)=M_{\mu,\nu}(2x\sqrt{-\lambda}), \hspace{1truecm}
\psi(x,\lambda)=W_{\mu,\nu}(2x\sqrt{-\lambda})$$ where the indices are 
$\mu=\frac{\alpha}{\sqrt{-\lambda}}$ and $\nu=\sqrt{\kappa +\frac 14}$. 
Their Wronskian 
\footnote{Since our study is effected in the sector of $|\Im \lambda | \leq 
\frac{1}{\epsilon}(\Re \lambda + \epsilon)$ then we assume that 
$\nu-\gamma >-\frac12$}  is 
$$ W=\frac{2\sqrt{-\lambda}\Gamma(2\nu +2)}{\Gamma(\nu-\mu+1)}.$$
The confluent hypergeometric functions coincide for $\alpha =0 $ 
with the Bessel functions:
$$M_{0,\nu}(x\sqrt{-\lambda})= 2^{2\nu}\Gamma(\nu+1)\sqrt{x}
I_{\nu}(\frac{x}{2}\sqrt{-\lambda}), \hspace{1truecm} 
W_{0,\nu}(x \sqrt{-\lambda})=\sqrt{\frac{x}{\pi}}
K_{\nu}(\frac x2\sqrt{-\lambda}),$$ the resolvent of $H_{\kappa,\alpha}$ 
could be obtained from that of $H_{\kappa}$ by Neumann series. The latter 
will be used in the operator estimates that follow. Therefore we form the 
Whittaker Green's function:
\begin{eqnarray*}
 R_{\lambda}(x,y)&=&\Theta(x-y)
\frac{\Gamma(\nu-\mu+\frac12)\Gamma(\nu-\mu+1)
M_{\mu,\nu}(2y\sqrt{-\lambda})W_{-\mu,\nu}(2x\sqrt{-\lambda})}
{2\sqrt{-\lambda}\Gamma(2\nu +2)\Gamma(\nu+\mu+\frac12)}\\ &+& 
\Theta(y-x)\frac{\Gamma(\nu-\mu+\frac12)\Gamma(\nu-\mu+1)
M_{\mu,\nu}(2y\sqrt{-\lambda})W_{-\mu,\nu}(2x\sqrt{-\lambda})}
{2\sqrt{-\lambda}\Gamma(2\nu +2)\Gamma(\nu+\mu+\frac12)}
\end{eqnarray*}

\subsection{Operator estimates.}
We denote by $R_{\lambda}^0=(\lambda-H_{\kappa})^{-1}$ the resolvent of 
$H_{\kappa}$. The positivity of $H_{\kappa}$ implies that 

\noindent {\bf 1.} The operator norm is 
$$ \| R^0_{\lambda} \|_{L^2} = O(|\lambda|^{-1})$$ for
$|\lambda| \rightarrow \infty$ uniformly in the cone  
$|\Im \lambda|\geq \frac{1}{\epsilon}(\Re \lambda+\epsilon)$:

\vspace{0.25truecm}
\noindent Since
$R^0_{\lambda}L^{2}(\overline{{\bf R}_{+}}) \subset {\cal D}(H_{\kappa})$ we
choose then $\phi \in L^{2}(\overline{{\bf R}_{+}})$:
 $R^{0}_{\lambda}\phi = \psi $. It follows that for $k \leq -\epsilon$:  
$$\|(\lambda-H_{\kappa})\psi \|^2_{L^2}=
|\lambda-k|^2\|\psi \|^2_{L^2} +
\|(k-H_{\kappa})\psi \|^2_{L^2} + 2(\Re(k-\lambda)(\psi,H_{\kappa}\psi)) $$
$$ \geq |(\lambda-k)|^{2}\|\psi \|^2_{L^2} $$
and the estimate follows.

\vspace{0.25truecm}
\noindent
The resolvent $R^{0}_{\lambda}$ is represented by the  
kernel for $\sigma=\frac{1}{\sqrt{-\lambda}}$:
$$R_{\sigma}(x,y)=(xy)^{1/2} 
[\Theta(x-y)K_{\nu}(\frac{x}{\sigma})I_{\nu}(\frac{y}{\sigma})
+\Theta(y-x)I_{\nu}(\frac{x}{\sigma})K_{\nu}(\frac{y}{\sigma})].$$  
We have the following classical estimates for the classical trace norms
$|| A ||_{k}:= tr(|A|^k)^{1/k}$:

\vspace{0.5truecm}
\noindent
{\bf 2.} Let $\phi\in C^{\infty}_{0}({\bf R})$ then 
$|| R_{\lambda}^{0} \phi||_{1} =O(|\lambda|^{-1/2})$ for $\lambda$ in the  cone 
$|\Im \lambda|\geq \frac{1}{\epsilon}(\Re \lambda+\epsilon)$.

\vspace{0.25truecm}
\noindent
Proof: For $ \lambda \leq -\epsilon $ and $\nu $ in the above cone it holds
that $$|| R_{\nu}^{0} \phi||_{1} \leq 
||(1+ (\lambda-\nu)R_{\lambda}^{0}) ||^{-1}_{L^{2}}
|| R_{\lambda}^{0}\phi ||_{1} $$ as well as that for
$\sigma^2=-\frac{1}{\lambda})$ :
$$|| R_{\lambda}^{0} \phi||_{1} = \int_{0}^{\infty}dx x\phi(x) 
(K_{\nu}I_{\nu})(\frac{x}{\sigma}). $$ Hence  we have the required
estimate for 
$$\int_{0}^{\infty}dx x\phi(x) (K_{\nu}I_{\nu})(\frac{x}{\sigma}) 
\leq \sigma||x\phi||_{L^{2}}||K_{\nu}I_{\nu}||_{L^{2}}.$$ 

\vspace{0.5truecm}
\noindent
{\bf 3.} Let $ \alpha \in \overline{{\bf R}_-},
\alpha + d \leq 2 $ then $$ \| x^{-\alpha}\partial_{x}^{d} 
R_{\lambda}^0 \|_{L^2} = O(|\lambda|^{\frac{\alpha +d}{2} -1})$$

\vspace{0.25truecm}
\noindent
Proof. This follows for $d=0$ from the inequality given in the beginning 
while for $d=2$ it results from the fact that 
$-\partial^{2}_x  \leq H_{\kappa + \frac14} $ and therefore
$$\| \partial_{x}^2 \phi \| \leq \| H_{\kappa +\frac14}R_{\lambda}^0\phi \|^2
\leq (\|\phi\|^2_{L^2}+|\lambda|^2\|R^0_{\lambda}\phi\|_{L^2}^2) .$$
Furthermore for $d=1$ we have for $\psi=R_{\lambda}^0\phi $ that
$$(x^{-\alpha}\partial_x\psi,x^{-\alpha}\partial_x\psi) =
(x^{-2\alpha}\psi,-\partial_x^2\psi) +
2\alpha(\partial_x\psi,x^{2(\alpha-\frac12}\psi) \leq$$
$$\leq 2\delta\|x^{2\alpha}\psi\|_{L^2}^2 + 2\delta^{-1}
\|\partial_x^2\psi\|_{L^2}^2 + 2\epsilon \| x^{-(2\alpha+1)}\psi \|^2_{L^2}
+2\epsilon^{-1}\| \partial_x \psi \|^2_{L^2} $$
the latter choosing $\delta= |\lambda|^{\alpha-\frac12}$ and 
$\epsilon=|\lambda|^{\frac{\alpha}{2}}$ gives the required estimate.

\subsection{ The Neumann series.} Let
 $\alpha > -2, k_0 =0 $  and $\{ k_n \}_{n=1}^{\infty}$ an 
increasing sequence of positive real numbers.
The potential $v$ has the asymptotic
expansion at the origin of the form: 
$$v(x) \sim x^{\alpha}\sum_{n=0}^{\infty}v_{n}x^{k_{n}}.$$  
In the sequel we will distinguish two cases:

\begin{itemize}

\item {\bf Case I.} Let  $\alpha =-1$ and  $ v_0 \neq 0$.

\noindent In this case the distributional trace of the resolvent 
$R_{\lambda}(H)$, $f_{H}(\frac{1}{\sqrt{-\lambda}}):=Tr(R_{\lambda}(H))$ 
is computed using Neumann series around 
$H_{\kappa,\alpha}=-\partial_x^2 +\frac{\kappa}{x^2}+
\frac{\alpha}{x}$:
$$f_{H}(\frac{1}{\sqrt{-\lambda}})=
\sum_{j=0}^{N}Tr(R_{\lambda}(H_{\kappa,\alpha})(v-\frac{\alpha}{x}))^j
R_{\lambda}(H_{\kappa,\alpha}) + Tr(R_{\lambda}(H_{\kappa,\alpha})
(v-\frac{\alpha}{x}))^{N+1}R_{\lambda}(H).$$
 
\item{\bf Case II.} Let  $\alpha > -1$ and $ v_0 \neq 0$. 

\noindent The Neumann series is based on the resolvent of
the usual Bessel operator $H_{\kappa}$. 
The case of $-2< \alpha <-1$ is treated within the frame of case B.  
\end{itemize}   

The Neumann series around the Bessel 
operator, due to its behaviour under scaling, is used in order to obtain  
the general form of the asymptotic expansion.

\section{The asymptotic expansion}

In this paragraph we 'll establish the existence and the precise form of the
asymptotic expansion of
the distributional trace of the resolvent $R_{\lambda}(H)$ of the operator 
$ H= -\partial_{x}^{2} + \frac{\kappa}{x^2} + v(x)$
by means of a Neumann series based on the resolvent $R_{\lambda}(H_{\kappa}$ 
of $ H_{\kappa}= -\partial_{x}^{2} + \frac{\kappa}{x^2} $, for a cut off
function $\chi \in C^{\infty}_{0}({\bf R}_{+}),\chi\equiv 1$ in a
neighbourhood of zero :
$$f_{\chi,H}(\frac{1}{\sqrt{-\lambda}}):=
Tr(\chi R_{\lambda}(H))=\sum_{j=1}^NTr(\chi(R_{\lambda}(H_{\kappa})v)^j
R_{\lambda}(H_{\kappa})) + O(|\lambda|^{\frac{N(\alpha-2)}{2}})$$
Denote by 
$I_{j}(\frac{1}{\sqrt{-\lambda}})=Tr(\chi(R_{\lambda}(H_{\kappa})v)^j
R_{\lambda}(H_{\kappa})).$ Actually we have the following 

\vspace{0.5truecm}
\noindent
{\bf Theorem.} {\it Let $\chi$ be a smooth cutoff function, $\chi \equiv 1$
in a neighborhood of the origin, then the function $f_{\chi,H}(\eta)$
has an asymptotic expansion of the form 
$$ f_{\chi,H}(\eta) \sim \sum_{k=0}^{\infty}A_{2k+1}\eta^{2k+1}+
\sum_{n=0}^{\infty}
B_{n}\eta^{z_n}+
\sum_{k=0}^{\infty}  C_{2k}\eta^{2k} +
 \sum_{k=0}^{\infty}  D_{2k+1}\eta^{2k+1}\log \eta $$
where $z_n \in {\bf R}\setminus {\bf Z},B_{k},C_{k},D_{k}$ are homogeneous 
polynomials in the asympotic coefficients $\{v_n\}$ of the potential 
of degree $z$ if we define $\mbox{deg}(v_n)=k_n+2$.} 

\vspace{0.5truecm}
\noindent
The proof will be achieved in two steps: first we prove the existence by 
appealing to the singular asymptotics lemma that we recall below and in the 
second step we use  a scaling argument in conjuction with  
the asymptotics of the resolvent of the Bessel operator 
$ H_{\kappa}= -\partial_{x}^{2} + \frac{\kappa}{x^2} $ which as it is well
known  it  is represented by the  kernel :
$$R_{\sigma}(x,y)=(xy)^{\frac12} 
[\Theta(x-y)K_{\nu}(\frac{x}{\sigma})I_{\nu}(\frac{y}{\sigma})
+\Theta(y-x)I_{\nu}(\frac{x}{\sigma})K_{\nu}(\frac{y}{\sigma})].$$

\vspace{0.5truecm}

\noindent {\bf Step 1.} The trace 
$$I_{j}= tr(\chi R_\lambda(H_\kappa)(vR_\lambda(H_\kappa))^{j-1})$$ 
is the $j+1$-tuple integral 
$$\int_0^\infty dx\int_0^\infty dx_1\dots \int_0^\infty dx_j
\chi(x)R_\lambda(x,x_1)\dots R_\lambda(x_j,x)$$ which is the sum of
integrals, each one for an ordering  of the the variables
$x,x_1,\dots,x_j$ hence for $x\geq x_1\geq \dots \geq x_j$:

$$\int_0^\infty dx \int_0^\infty dx_1 \dots \int_0^{x_{j-1}} dx_j
x^2\chi(x)x_1^2v(x_1)\dots
x_j^2v(x_j)(K_\nu(\frac{x}{\sigma})I_\nu(\frac{x_j}{\sigma}))^2
\prod_{i=1}^{j-1}K_\nu(\frac{x_i}{\sigma})I_\nu(\frac{x_i}{\sigma})$$ and we
perform the  coordinate change (blow up)
$$x_1=xt_1,\dots,x_j=x_{j-1}t_{j-1} $$ and
$$x_1=x\theta_1,\dots,x_j=x\theta_{j-1}, \theta_k=t_1\dots t_k$$
that leads to a sum of integrals of the form: 
$$\int_0^\infty dx x^{2j}\chi(x)
\int_0^1\frac{d\theta_1}{\theta_1}\dots\int_0^1\frac{d\theta_j}{\theta_j}
\prod_{k=1}^j v(x\theta_j)\theta_j^2
K^2_\nu(\frac{x}{\sigma})I^2_\nu(\frac{x}{\sigma}\theta_j)
\prod_{k=1}^{j-1}(K_\nu I_\nu)(\frac{x}{\sigma}\theta_k) $$
which is finally of the form suggested from the SAL

$$ {\cal I}_{j}(\sigma)= 
\int_{0}^{\infty} \frac{dx}{x} \chi (x)x^{3} \cdot  F(\xi,x)$$ 
where $$F(\xi,x):= \int_{0}^{1}\frac{d\theta_{1}}{\theta_{1}} \cdots 
\int_{0}^{1}  \frac{d\theta_{j}}{\theta_{j}} 
({\cal B}_{\nu}(\xi, \theta_{j})
{\cal R}_{j\nu}(\xi;\theta_{1}, \cdots, \theta_{j-1})
 v_{j}(x;\theta_{1}, \cdots, \theta_{j}). $$
if we  set 
$${\cal B}_{\nu}(\xi, \theta_{j}):=
(K_{\nu}(\frac{1}{\xi}) I_{\nu}(\frac{\theta_{j}}{\xi}))^{2} ,
\hspace{1truecm} {\cal R}^{j}_{\nu}(\xi,\theta_{1}, \cdots, \theta_{j-1})
:= \prod_{i=1}^{j-1}
(K_{\nu}I_{\nu})(\frac{\theta_{i}}{\xi})$$
and 
$$v_j(x;\theta_{1}, \cdots, \theta_{j}):= 
\prod_{i=1}^{j}(x\theta_{i})^2v(x\theta_{i})=O(x^{j(\alpha +2)}) $$
Now we see that $v_{j} \in \Gamma^{\infty}(({\bf R}_{+}))$ with asymptotic 
character $\underbrace{S + \cdots + S}_{j-times}$ since 
 $v \in \Gamma^{\infty}({\bf R}_{+})$ has character $S$, 
$v(x)=O(x^{\alpha}), \alpha > -2$. Furthemore the character of 
$({\cal Q}_{\nu} \cdot {\cal P}^{j}_{\nu})(\xi,\theta_{1}, 
\cdots, \theta_{j})$ is $S(k)=1$ for $k\in j+1+2{\bf Z}_+$. The estimates 
for the Bessel functions as $\xi \rightarrow 0$:
$$K_{\nu}(\frac{1}{\xi})I_{\nu}(\frac{\theta}{\xi})= 
O(\xi), \hspace{1.5truecm}
K_{\nu}(\frac{\theta}{\xi})I_{\nu}(\frac{\theta}{\xi})= 
O(\xi).$$
From these it  follows that $F(\xi,x) \in \Gamma^{\infty} (({\bf R}_{+})^{2})$
and also that 
$$ | {\overline P_{z}^{x}} F(\xi, x)| \leq (x\xi)^{\Re z + \delta}h_{z}(\xi)$$
where we have that 
$$ {\overline P_{z}^{x}} = \prod_{Re z'\leq Re z}(x\partial_{x} - z')^{S(z')}, 
\hspace{1truecm} \int_{1}^{\infty}h_{z}\frac{dz}{z} < \infty .$$  
Then we conclude that $${\cal I}_{j}(\sigma) \sim 
\sigma^3\sum_{n}^{\infty} 
B^{(j)}_{n}\sigma^{z_n} +
\sigma^{j+1}(\sum_{k=0}^{\infty} A^{(j)}_{k}\sigma^{2k}+ 
\sum_{k=0}^{\infty} C^{(j)}_k\sigma^{2k+1}
+\sum_{k=0}^{\infty} D^{(j)}_k\sigma^{2k}(\log \sigma)) $$
where the $ z_n \notin j+1+2{\bf Z}_+$ and the  
$B^{(j)}_{z},C^{(j)}_{k},D_k^{j}$ are polynomials in the coefficients of the 
potential of the potential of degree $k+j$ if we define $deg v_{z}= z+2$.
These suffice for the  $B'$s  in the expansion.

\vspace{0.5truecm}
\noindent
{\bf Step 2.} We have to establish the nature of the non classical terms
logarithmic terms. This will be achieved by induction on the ordered 
couples $(j,k)$ where $j$ is the order of the term in the Neumann series and
$k$ is order in the asymptotic expansion of the potential.   
For $k=\alpha$ the assertion follows from above: everything depends on 
$v_{0}$; indeed $j=0$  follows from \cite{c2}. Assume that $k > \alpha $. 
Then consider 
$$I_{j}= Tr(\chi (R_{\lambda}(H_{\kappa})v)^{j}R_{\lambda}(H_{\kappa})$$
Notice that up to $O(\lambda^{-\infty})$ we could embody the cutoff function
in the potentials $V$. Accordingly let $N \in {\bf Z}, k_N \geq -2$ and split 
the potential as $v=v_N + R_N$, where 
$v_N(x) \sim  x^{\alpha}\sum_{n=0}^{N}v_{n}x^{k_{n}}$.  
The following commutator identities for $k_r \geq \frac{k_n}{2}-1, r<n$ and
$\epsilon=k_n-k_r>0$ give that 
$$x^{k_{n}} R_{\lambda}(H_{\kappa}) =
x^{k_r}R_{\lambda}(H_{\kappa})x^{\epsilon}+2\epsilon x^{k_r}
R_{\lambda}(H_{\kappa})(x^{\epsilon-2}(x\partial_{x} +\frac{\epsilon-1}
{2\epsilon})R_{\lambda}(H_{\kappa}),$$ 
$$x\partial_x R_{\lambda}(H_{\kappa})=
 R_{\lambda}(H_{\kappa})x\partial_x -
2R_{\lambda}(H_{\kappa})^2H_{\kappa}$$ 
and in turn allow us  to vary the number of the $R_{\lambda}(H_{\kappa})$ 
factors. This in view of the  fact that  the asymptotic expansion of 
$I_{\nu}K_{\nu}(x)$ as $x \rightarrow \infty$ contains only odd powers,
in view of   $\lambda\partial_{\lambda}R_{\lambda}(H_0)=-
2\sigma\partial_{\sigma}R_{\sigma}$,  leading to  the polynomial dependence of
the coefficients of the logarithms and the even powers:
$$I_j(\eta) \sim \sum_{k=0}^{\infty}A_{2k+1}\eta^{-2k-1}+
\sum_{n=0}^{\infty}  B_{z_n}\eta^{-z_n}+
\sum_{k=0}^{\infty}  C_{2k}\eta^{-2k} +
 \sum_{k=0}^{\infty}  D_{k}\eta^{-k-4}\log \eta $$
where $z_n \in {\bf R} \setminus {\bf Z}, z >-2$  and the 
$B_{z},C_{k},D_{k}$ are homogeneous polynomials in the asympotic 
coefficients $\{v_{n} \}_{n=0}^{\infty}$ of $v$  
of degree $z$ if we define $deg(v_{z})=z+2$.  Then summing up we obtain an 
expansion of the same form and we appeal to 
the results in par. 5 of \cite{c1} in order to keep the odd terms in the log's.

\noindent In conclusion we supply the asymptotics of 
$$G_{t}(\xi):= K_{\nu}(\frac{1}{\xi})I_{\nu}(\frac{t}{\xi})$$ 
as $\xi \rightarrow 0$ uniformly in $0 \leq t \leq 1$. For this we 
calculate the Mellin transform using the integral representation:
$$G_t(\xi)=K_{\nu}(\frac{1}{\xi})I_{\nu}(\frac{t}{\xi})=
\int_{0}^{\infty} \frac{x^2}{x^2+\xi^{-2}}J_{\nu}(tx)J_{\nu}(x)\frac{dx}{x}.$$
The Hankel transforms \cite{t2} allows us to obtain further:
$$G_t(\xi)\frac{(2t)^{\nu}}{\sqrt{\pi}\Gamma(\nu+\frac12)}
\int_{0}^{\infty}
 \frac{x^{\nu+2}}{x^2+\xi^{-2}}\int_0^1\frac{J_{\nu}(x\sqrt{1+t^2-2ut})}
{(1+t^2-2ut)^{\frac{\nu}{2}}}du\frac{dx}{x}.$$
From this we get that the Mellin transform is 
$$\widehat{G}_t(s)=
-\frac{2^{-s+2\nu-2}\Gamma(\nu-\frac s2)\Gamma(-\frac s2)}
{\sqrt{\pi}\Gamma(\nu+\frac12)} (1+t)^{s+\nu}\int_0^1
\eta^{\nu-\frac12}(1-\eta)^{\nu-\frac12}
(1-\frac{2t}{(1+t)^2}\eta)^{s-\nu}d\eta.$$
A Taylor expansion then leads to the series
  $$\widehat{G}_t(s)=
-(1+t)^{s-2\nu}\frac{2^{-s+2\nu-2}\Gamma(\frac s2+\frac12)
\Gamma(\nu-\frac s2)\Gamma(-\frac s2)}
{\sqrt{\pi}\Gamma(\nu+\frac12)}
\sum_{k=0}^{\infty}\frac{(-1)^kI_k(t)}{k!(\Gamma(\frac s2+\frac12-k)}
\frac{(2t)^k}{(1+t)^{2k}} $$ where 
$$ I_k(t)=\int_0^1\eta^{\nu-\frac12+k}
(\frac{1-\eta}{1-\frac{2t}{(1+t)^2}\eta})^{\nu-\frac12}d\eta.$$
This integral satisfies that $I_k = O(k^{-\frac12 -\delta})$ as $k \rightarrow\infty$, 
uniformly in $t$, where $\delta=\frac12$ for $1>\epsilon >0, \nu \geq \frac12$
and $\delta=\nu$ for $\nu<\frac 12$. Stirling' s formula then suggests that 
the summand behaves as $O(k^{-\frac s2-1-\delta})$ and hence it is uniformly convergent 
for $0<t<1$.  Finally we  conclude that the asymptotic expansion of  
$g_t(\xi)$ as $\xi \rightarrow 0$  contains only odd powers of $\xi$.
\footnote{In order to extend meromorphically  it out of the strip $-1+\nu < \Re s < \nu$
 we differentiate with respect to $\zeta=\frac{2t}{(1+t)^2}$. The poles
as $t \rightarrow 1 $ remain located there. }

\section{The recursions for the coefficients of the expansion} 

We will consider the two cases indicated in section 2.4 as follows:

\subsection{Case I.}

The first order term in the Neumann series is reduced
$mod(\lambda^{-\infty}),$ to
$$I_1(\frac{1}{\sqrt{-\lambda}})=
tr(R_{\lambda}(H_{\kappa,v_0})(v-\frac{v_0}{x})R_{\lambda}(H_{\kappa,v_0})\chi)=
-\partial_{\lambda}tr(R_{\lambda}(H_{\kappa,v_0})(v-\frac{v_0}{x})\chi)$$
Setting $\sigma=\frac{1}{\sqrt{-\lambda}}$ we arrive through the formula  
$$I_1(\sigma)=\frac{\sigma^2}{2} \sigma\partial_{\sigma}\tilde{I_1}(\sigma)$$ 
at the integral
\begin{equation}
\tilde{ I_1}(\sigma)=  \sigma \cdot 
\int_{0}^{\infty}{\cal G}_{\mu,\nu}(\frac2y)V(x)\frac{dx}{x}
\end{equation} where 
$$V(x)= x(v(x)-\frac{v_0}{x}), \hspace{1truecm} {\cal G}_{\mu,\nu}(\frac2y)=
\frac{\Gamma(\nu-\mu+\frac12)}{2\Gamma(2\nu +1)} 
M_{\mu,\nu}W_{\mu,\nu}(\frac2y), $$
for $M_{\mu,\nu},W_{\mu,\nu}$ being the Whittaker functions
$y=\frac{\sigma}{x}, \mu= \frac{v_0xy}{2}$. This Green's function could
be simplified in the form that we 'll use in the sequel
$${\cal G}_{\mu,\nu}(y):=\frac{{\tilde{\cal G}}_{\mu,\nu}(y)}
{\alpha_{\nu}(\mu)}$$ where 
$${\tilde{\cal G}}_{\mu,\nu}(y)=y^{2\nu+1}{\tilde W}_{\mu,\nu}(y)
{\tilde M}_{\mu,\nu}(y) $$ for the integrals
$${\tilde M}_{\mu,\nu}(y)=\int_{-1}^{1}e^{-y(1-\theta_1)}
(1-\theta_1^2)^{\nu-\frac12}(\frac{1-\theta_1}{1+\theta_1})^{\mu}d\theta_1,$$
$${\tilde W}_{\mu,\nu}(y)=\int_0^{\infty}e^{-2y\theta_2}
(\theta_2(1+\theta_2)^{\nu-\frac12}(\frac{1+\theta_2}{\theta_2})^{-\mu}
d\theta_2,$$ $$\alpha_{\nu}(\mu)=\Gamma(\nu-\mu+\frac12)
\Gamma(\nu+\mu+\frac12).$$

\noindent The study of the $\sigma \rightarrow 0$ asymptotics of the above intgeral
will provide the inverse spectral result. Indeed the Singular Asymptotics 
lemma provides the asymptotic coefficients of the integral 
$\sigma \rightarrow 0$. For brevity we introduce the notation
$${\cal C}_{j,\nu}(y)=\partial^j_{\mu}
{\cal G}_{\mu,\nu}(\frac2y)|_{\mu=0}, \hspace{0.5truecm}
c_{j,\nu}(y)=\partial^j_{\mu}
{\tilde {\cal G}}_{\mu,\nu}(\frac2y)|_{\mu=0}.$$ 
Then we distinguish between  the following cases:

A. $2\nu\in {\bf Z}_+, \nu=n+\frac12$. For $l=0,\dots$ we
appeal to the logarithmic terms 
\begin{eqnarray*}
D_{2l+1,1}\tilde{I_{1}}(0) &=&  -\frac{1}{(2l)!}
D_{2l,0}^{y}\partial_{x}^{2l}|_{x=0}[{\cal G}_{\mu,\nu}(\frac2y)
\cdot V(x)]\\
&=&  -\sum_{j=0}^{2l}(\frac{v_{0}}{2})^j
[(D_{2l,0}^{y}(y^{j}{\cal C}_{j,\mu}(y))]v_{2l-j-1}
\end{eqnarray*}
and also to the even powers that are given 
$$ D_{2l,0}{\tilde I_{1}}(0) = 
-\sum_{j=0}^{2l}(\frac{v_{0}^j}{2^jj!})
[u^y_{2l,0}(y^j {\cal C}_{j,\mu}(y))]v_{2l-j-1}.$$
Actually for the inverse spectral result we'd like the explicit forms: 
\begin{eqnarray*}
D_{2l+3,1}\tilde{I_{1}}(0)&=&
B_{2l+2}^0(\nu)v_{2l+2}
-(l+1)B_{2l+2}^1(\nu)v_{0}v_{2l+1}-(l+1)(l+\frac12)
B_{2l+2}^2(\nu)v_{0}^{2}v_{2l}\\
&+&P_{2l+2}(v)
\end{eqnarray*}
and also that
\begin{eqnarray*}
D_{2n+2j+2,0}\tilde{I_{1}}(0)&=&
C_{2j+2n+1}^0(\nu)v_{2n+2j+1}+(j+n+\frac12)C_{2j+2n+1}^1(\nu)v_{0}v_{2n+2j}\\
&+&Q_{2n+2j+1}(v)
\end{eqnarray*}
where we have denoted the coefficients 
$$B_{k}^j(n):=D_{k,0}^{y}(y^j {\cal C}_{j,\nu})=D_{k-j,0}{\cal C}_{j,\nu},
\hspace{0.5truecm} 
 C_{k}^j(\nu):=u_{k,0}^y(y^j{\cal C}_{j,\nu})=u_{k-j,0}({\cal C}_{j,\nu})$$
and the $P,Q$ are the polynomials suggested precedingly by the general form
of the asymptotic expansion. Notice that if  $l=0,\dots,n $ then we 'll
see in the sequel that: 
$$B^0_{2l}(\nu)\neq 0, \hspace{0.5truecm}  C^0_{2l+1}(\nu)\neq 0$$
 therefore coefficients $v_k, k=1,\dots,2n+1$ are
determined immediately. For the remaining we appeal to 
the pair of equations; hence we have to establish, provided that 
$v_0 \neq 0$, that the determinant: 
$$ \Delta_{k}(\nu)=B_{2n+2k+2}^1(\nu)C_{2n+2k+1}^1(\nu)-
B_{2n+2k+2}^2(\nu)C_{2n+2k+1}^0(\nu)$$
does not vanish.

B. $2\nu \in  {\bf R \setminus Z_+}.$  The inverse spectral
result requires again the logarithmic odd powers as above, which are
provided from the first order terms 
$$D_{2l+3,1}\tilde{I_1}(0)=D_{2l+3}^{(0)}(\nu)v_{2l+1}+P_{2l+2}(V)$$ as well
as the the pure powers $\alpha \in {\bf R}, \alpha > -2, \alpha\neq
0,2,\dots$:
  $$ D_{\alpha,0}{\tilde I}_1(0)= u_{\alpha,0}^{y}[({\bar P}_{\alpha}
{\cal G}_{\mu,\nu}(\frac2y) \cdot V]|_{x=0}.$$ In this formula  
${\bar P}_{\alpha}=\prod_{z \leq \alpha}(x^z x\partial_x x^{-z})$ and hence
we obtain the set of equations 
$$D_{\alpha+2,0}\tilde{I_1}(0)= C_{\alpha+2}^0(\nu)v_{\alpha+1}+
C_{\alpha+2}^1(\nu)v_{0}v_{\alpha+1}+  C_{\alpha+2}^2(\nu)v^2_{0}v_{\alpha}+  
Q_{\alpha+2}(v).$$ Notice that if $\alpha \neq 2\nu+2j+1$ then 
$C_{\alpha+2}^0\neq 0$ and hence $ v_{\alpha} $ is determined by the first 
terms for $\alpha \neq 2\nu+k$. These in particular contain the coefficients 
for $\alpha=2\nu+2j+1, j=0,\dots.$
\begin{eqnarray*} 
D_{2\nu+2l+4,0}{\tilde I}_{1}(0)&=&C_{2\nu+2l+3}^{1}(\nu)v_{0}v_{2\nu+2l+2}+  
C_{2\nu+2l+3}^2(\nu)v_0^2v_{2\nu+2l+1}+ Q_{2\nu+2l+3}(v),\\
D_{2\nu+2l+3,0}\tilde{I_1}(0)&=&C_{2\nu+2l+2}^0(\nu)v_{2\nu+2l+2}+
C_{2\nu+2l+2}^1(\nu)v_{0}v_{2\nu+2l+1}+Q_{2\nu+2l+2}(v)
\end{eqnarray*}
where $Q$ are the same polynomials that appear also above. Again we have to
stablish that  determinant: 
$$\Delta_{k}(\nu)=C_{2\nu+2k+2}^1(\nu)C_{2\nu+2k+1}^1(\nu)-
C_{2\nu+2k+2}^2(\nu)C_{2\nu+2k+1}^0(\nu)$$ 
of the coefficients of the preceding system of equations is nonvanishing.

\vspace{0.25truecm} \paragraph{The Mellin Transforms of $C_{j,\nu}$}
Finally all these asymptotic coefficients coincide with the coefficients of
the Laurent expansion of the Mellin transform $$\hat{c}_{j,\nu}(s)=
\int_{0}^{\infty}x^{s}c_{j,\nu}(x)\frac{dx}{x}$$ as it is explained
in the appendix on the Singular Asymptotics Lemma. In the sequel we' ll
employ the identification provided by the identities: 
\begin{equation}
D_{k,0}(c_{j,\nu})(0)= Res_{s=-k}\hat{c}_{j,\nu}
\hspace{1truecm} u_{k,0}(c_{j,\nu})=\hat{c}_{j,\nu}(-k)
\end{equation}

\noindent The function $c_{j,\nu}(y)$ is written under the following 
change of  variables 
$$\theta_1=2\overline{\theta}_1-1, \hspace{1truecm}\theta_2=
\frac{(1-\overline{\theta_1})\overline{\theta_2} }{1-\overline{\theta}_2}$$
in the form: $$c_{j,\nu}(y)= \int_{0}^{1}\int_0^1
\frac{d\theta_1d\theta_2}{1-\theta_2}
(\theta_1\theta_2)^{\nu-\frac12}(1-\theta_1\theta_2)^{\nu-\frac12}
(\frac{1-\theta_1}{1-\theta_2})^{2\nu-1}\log^j(\frac{1-\theta_1\theta_2}
{\theta_1\theta_2})\exp(-\frac 2y\frac{1-\theta_1}{1-\theta_2}).$$ Further
we perform the change of variables $$ (0,1)\times (0,1)
\ni(\theta_1,\theta_2) \mapsto (\xi,\eta) \in (0,1) \times (0,\infty),$$
$$\xi=\theta_1\theta_2, \hspace{1truecm}
\eta=\frac{1-\theta_1}{1-\theta_2},$$
$$\frac{d\theta_1 d\theta_2}{1-\theta_2}=\frac{d\xi d\eta}{2r(\xi,\eta)
(1+\eta)}$$ where
  $r(\xi,\eta)=(1-\frac{4\xi\eta}{(1+\eta)^2})^{\frac12}.$
In these coordinates the preceding 
integral is written as: $$c_{j,\nu}(y)=\int_0^{\infty}
\frac{\eta^{2\nu+1}e^{-\frac{2\eta}{y}} d\eta}{1+\eta} 
\int_0^1\frac{d\xi}{r(\xi,\eta)}[\xi(1-\xi)]^{\nu-\frac12}
\log^j(\frac{\xi}{1-\xi}).$$ In the sequel we 'll derive its asymptotics
as $y\rightarrow 0$. To that end,  we perform the change of variable
${\tilde \xi}=\frac{\xi}{1-\xi}$ to obtain finally by abusing notation
for the function $r(\xi,\eta)$ that
$$c_{j,\nu}(y)=\int_0^{\infty} 
\frac{e^{-\frac{2\eta}{y}}\eta^{2\nu+1}d\eta}{1+\eta}
\int_0^{\infty}\frac{\xi^{\nu-\frac12}d\xi}{r(\xi,\eta)}
\frac{\log^j\xi}{(1+\xi)^{2\nu+1}}.$$

\noindent By a Taylor expansion of the function $r(\xi,\eta)$  we obtain the
Mellin transform of the function $c_{j,\nu}(s)$:
$$\widehat{c}_{j,\nu}(s)=(-1)^j2^{s-2\nu-1}\Gamma(-s+2\nu+1)
\sum_{l=0}^{\infty}d_l^j(\nu)\Gamma(s+n+1)\Gamma(n-s) $$ where 
$$d_l^0(\nu)=\frac{(2l+1)B(l+\nu+\frac12,\nu+\frac12)}{(\Gamma(l+1))^2}$$
$$d_l^1(\nu)=d_l^0(\nu)[\psi(l+\nu+\frac12)-\psi(\nu+\frac12)],$$
$$d_l^2(\nu)=d_l^0(\nu)[(\psi(l+\nu+\frac12)-\psi(\nu+\frac12))^2-
\psi'(l+\nu+\frac12)-\psi'(\nu+\frac12)]$$

\noindent These combined with Stirilng's formula suggests that the series
converges absolutely for $0< \Re  s <1$ and represents a meromorphic 
function with poles at the integral points. Using the Laurent expansion of
the Gamma function we obtain that for $l<k$
\begin{eqnarray*}
&& 2^{s-1}\Gamma(-s+2\nu+1)\Gamma(s+l+1)\Gamma(l-s)=\\
&&\frac{(-1)^{k-l}\Gamma(l+k+1)\Gamma(k+2\nu+1)}{2^{k-1}\Gamma(k-l)}\cdot\\
&&[\frac{1}{s+k}-(\psi(l+k+1)+\psi(k-l+1)-\log2+\psi(2\nu+k+1)+O(s+k)]
\end{eqnarray*}
and for $l\geq k$
$$2^{s-1}\Gamma(-s+2\nu+1)\Gamma(s+l+1)\Gamma(l-s)
=2^{-(k+1)}\Gamma(k+2\nu+1)
\Gamma(l-k+1)\Gamma(l+k+1) +O(s+k).$$
Furthermore let $N=[2\nu]$ then the Taylor expansions at  $-(2\nu+k), k\in
{\bf Z}_+$ give that for $l\leq N+k$:
$$ 2^{s-1}\Gamma(-s+2\nu+1)\Gamma(s+l+1)\cdot\Gamma(l-s)=
\frac{\Gamma(k+4\nu+1)\Gamma(2\nu+k+l+1)}{\sin2\nu\pi\Gamma(2\nu+k-l)}
 +O(s+2\nu+k)$$
and for $l\leq N+k$ we have 
$$ 2^{s-1}\Gamma(-s+2\nu+1)\Gamma(s+l+1)\cdot\Gamma(l-s)=
2^{-(2\nu+k+1)}\Gamma(k+4\nu+1)\Gamma(l+1-2\nu+k)\Gamma(2\nu+k+l)
+O(s+2\nu+k)$$
            
\noindent {\bf The identities.}
At this point notice that since $${\tilde M}_{0,\nu}(x)=2^{\nu}
\Gamma(\nu+\frac12)\sqrt{\pi}x^{-\nu}e^{-x} I_{\nu}(x), \hspace{0.5truecm} 
{\tilde W}_{0,\nu}(x)=\frac{2^{-\nu}\Gamma(\nu+\frac12)}{\sqrt{\pi}}
x^{-\nu}e^xK_{\nu}(x)$$ then $${\cal C}_{0,\nu}(y)=yI_{\nu}(y)K_{\nu}(y)$$
and this suggests further that $${\widehat {\cal C}}_{0,\nu}(s)=
\frac{\Gamma(\frac12-\frac s2+\nu)\Gamma(\frac12-\frac s2)
\Gamma(\frac s2)} {4\sqrt{\pi}\Gamma(\frac12+\nu+\frac s2)}.$$ 
The general form of the asymptotic expansion suggests the absence of
logarithms  in the odd powers that in turn implies the identities  
$B_{2m+1-j}^j(\nu)=0$. In parallel the exact form of 
$\widehat{{\cal C}}_{0,\nu}$ suggests that for $\nu=n+\frac12$ then 
$B_{k}^0(n+\frac12)=0$ for $k=-2(n+l)$ whereas $C_{2\nu+2k+3}^0(\nu)=0$. 
We 'll employ later these identities but prior to that we study the resulting
coefficients.

\noindent {\bf The asymptotic coefficients in the recurrence relation.} We
treat the preceding cases separately.

\smallskip

\noindent {\bf A. $\nu=n+\frac12$.} The required coefficients are given by

$$B^j_{k+j+1}(\nu)=\frac{(-1)^k\Gamma(2n+k+j+3)}{2^{2n+k+j+1}}[
\beta_{k+j,0}^j(\nu)-\beta_{k+j,1}^j(\nu)]$$
where for $j=1,2$
\begin{eqnarray*}
&&\beta_{2k+1,0}^1(\nu)=\sum_{l=0}^kd^j_{2l}(\nu)
\frac{\Gamma(2k+2+2l)}{\Gamma(2k+2-2l)} \\
&&\beta_{2k+1,1}^1(\nu)=\sum_{l=0}^kd^j_{2l+1}(\nu)
\frac{\Gamma(2k+2l+3)}{\Gamma(2k+1-2l)} \\
&&\beta_{2k+2,0}^2(\nu)=\sum_{l=0}^{k+1}d^j_{2l}(\nu)
\frac{\Gamma(2k+3+2l)}{\Gamma(2k+3-2l)} \\
&&\beta_{2k+2,1}^2(\nu)=\sum_{l=0}^{k+1}d^j_{2l+1}(\nu)
\frac{\Gamma(2k+2l+4)}{\Gamma(2k+3-2l)} 
\end{eqnarray*} 

\noindent as well as that for $\delta_{ij}$ the usual Kronecker symbol:
\begin{eqnarray*}
&&C^j_{k+j+1}(\nu)= -\frac{(-1)^k\Gamma(k+2n+j+2)}{2^{k+j+2n+3}}[
\gamma^j_{k+j+1,0}(\nu)-\gamma^j_{k+j+1,1}(\nu)\\
&&+(1-\delta_{j0})[\gamma_{k+j+1,0}^0(\nu)-\gamma^j_{k+j+1,1}(\nu)]
+(\psi(k+j+2n+3)-\log2)+r_{k+j+1}(\nu)]
\end{eqnarray*}

where now for $j=0,1$

\begin{eqnarray*}
&& \gamma_{2k+2,0}^0(\nu)=\sum_{l=0}^{k+1}d^j_{2l}(\nu)
\frac{\Gamma(2k+2l+4)}{\Gamma(2k+3-2l)}\bigl(\psi(2l+2k+3)+
\psi(2k-2l+3)\bigr)\\
&&\gamma_{2k+2,1}^0(\nu)=\sum_{l=0}^{k+1}d^j_{2l+1}(\nu)
\frac{\Gamma(2k+2l+4)}{\Gamma(2k+3-2l)}\bigl(\psi(2l+2k+4)+
\psi(2k+3-2l)\bigr)\\
&&\gamma_{2k+1,0}^1(\nu)=\sum_{l=0}^kd^j_{2l}(\nu)
\frac{\Gamma(2k+2l+3)}{\Gamma(2k+2-2l)}\bigl(\psi(2l+2k+2)+
\psi(2k-2l+2)\bigr)\\
&&\gamma_{2k+1,1}^0(\nu)=\sum_{l=0}^kd^j_{2l+1}(\nu)
\frac{\Gamma(2k+2l+3)}{\Gamma(2k+2-2l)}\bigl(\psi(2l+2k+3)+
\psi(2k+2-2l)\bigr)\\
&&r_{m}(\nu)= \sum_{l=0}^\infty d_{l+m}^j(\nu)
\Gamma(l+1)\Gamma(l+2m+1)
\end{eqnarray*}

We will use the above in order to get
$B^1_{2n+2k+2}(\nu),B^2_{2n+2k+2}(\nu),
C_{2n+2k+1}^0(\nu),C_{2n+2k+1}^1(\nu)$.

\noindent {\bf II. $2\nu\notin {\bf Z}_+$.} Let $N=[2\nu]$ then we get that :

$$C^j_{2\nu+k+j}(\nu)= 
\frac{\Gamma(k+4\nu+j+1)}{2^{4\nu+k+j+1}}][
\frac{\pi}{\sin2\nu\pi}(-1)^k\bigl(
\gamma_{N+k+j,0}^j(\nu)-\gamma_{N+k+j,1}^j(\nu)\bigr)
+\rho_{N+k+j+1}(\nu)]$$

where

\begin{eqnarray*}
&& \gamma_{N+k+j,0}^j(\nu)=
\sum_{l=0}^{[\frac{N+k+j}{2}]}d_{2l}^j(\nu)
\frac{\Gamma(2\nu+k+2l+j+2)}{\Gamma(2\nu+k+j-2l)}\\
&&\gamma_{N+k+j,1}^j(\nu)=
\sum_{l=0}^{[\frac{N+k+j}{2}]}d_{2l+1}^j(\nu)
\frac{\Gamma(2\nu+k+2l+j+3)}{\Gamma(2\nu+k+j-2l-1)}\\
&&\rho_{N+k+j+1}(\nu)=
\sum_{l=0}^{\infty}d_{l+N+k+j+2}^j(\nu)
\Gamma(l+N-2\nu+1)\Gamma(l+2\nu+N+2k+2j+1)
\end{eqnarray*}

\noindent The preceding formulae will provide
$C^0_{2\nu+2k+2}(\nu),C^1_{2\nu+2k+1}(\nu),
C_{2n+2k+2}^1(\nu),C_{2n+2k+2}^2(\nu)$.

\paragraph{Certain elementary estimates.} An application of the steepest
descent
provides the following elementary estimate for the $\Gamma$-function,
for $\delta\in (0,\frac12),x\geq 1$:
$$c_\delta(x+1)^xe^{-(1-\frac{4\delta^2}{2-\delta^2})}\leq
\Gamma(x)\leq C_\delta(x+1)^xe^{-(1-\delta^2)x}$$
while the Euler-MacLaurin formula gives for the $\psi,\psi'$-functions 
and $a>0,k\in {\bf Z}_+$:
$$\pi_{21}(a)<\psi(k+a+1)-\log(k+1)+\frac{1}{6(k+a)}<\pi_{11}(a),$$
$$\pi_{12}(a) <\psi'(k+a+1)-\frac{k-1}{(a+1)(k+a)}+\frac{1}{3(k+a)^2}<
\pi_{22}(a)$$ as well as that 
$$\pi_{11}(a)=\psi(a)+\frac1a+\frac{4a+3}{6(a+1)^2},
\pi_{21}(a)=\psi(a)+\frac1a+\frac{2a+3}{6(a+1)^2},$$
$$\pi_{12}(a)=\psi'(a)-\frac{1}{a^2}+\frac{3a+5}{6(a+1)^3}+
\frac{a+6}{8(a+1)},$$
$$\pi_{22}(a)=\psi'(a)-\frac{1}{a^2}+\frac{3a+5}{6(a+1)^3}-
\frac{a+6}{8(a+1)},$$
Actually we have the elementary inequality for $t>0$, $$min(a,b)\leq
\frac{e^{at}-1}{e^{bt}-1}\leq max(a,b)$$ which is used after the 
$$n^{-s}=\frac{1}{\Gamma(s)}\int_0^\infty t^s e^{-nt}\frac{dt}{t}$$
These imply that for $\beta_\delta<1$:
$$ \frac{\kappa_1(\delta,\nu)\beta_\delta^l}{(l+1)^{2l+\nu+\frac12}}
\leq d_l^0(\nu)
\leq \frac{\kappa_2(\delta,\nu)}{\beta_\delta^l(l+1)^{2l+\nu+\frac12}}.$$
that allow us to obtain that for 
$\zeta_\delta(m)=2m+\delta^2+3,
\eta_1(\delta)=\frac{2\delta^2-15\delta-8}{2-\delta^2},
\eta_2(\delta)=\frac{2\delta^2-7\delta+14}{2-\delta^2}$ the estimates

\begin{eqnarray*}
&&\kappa_1(\delta)<\beta_{m,l}^0(\nu)\leq
\frac{\kappa_2(\delta,\nu) e^{m\zeta_\delta(m)}}{m^{2\nu}
\zeta_\delta(m)^{\nu+\frac12}}
(1+O(\zeta^{\nu+\frac12}m^{2\nu}e^{-\zeta_\delta(m)m})),\\
&& \kappa_1(\nu)<\gamma_{m,l}^0(\nu)\leq
\frac{\kappa_2(\delta,\nu) e^{m\zeta_\delta(m)} }{m^{2\nu}
\zeta_\delta(m)^{\nu+\frac12}}
(1+O(\zeta^{\nu+\frac12}m^{2\nu}e^{-\zeta_\delta(m)m}))\\
&&\frac{\kappa_2(\delta,\nu)
e^{(\log4-2-\eta_2(\delta))m}}{(1+m)^{\nu+\frac52}}
<r_m(\nu),\rho_m(\nu)<
\frac{\kappa_1(\delta,\nu)e^{(2+\log4-\eta_1(\delta))m}}{(1+m)^{\nu+\frac52}}
\end{eqnarray*}
At this point we remark that the preceding identities are expressed in this
notation in the form
\begin{eqnarray*}
&& \beta_{2k+1,0}^0(n+\frac12)=\beta_{2k+1,1}^0(n+\frac12),\\
&&\frac{\pi}{\sin2\nu\pi}(-1)^k\bigl(
\gamma_{N+k+j,0}^0(\nu)-\gamma_{N+k+j,1}^0(\nu)\bigr)
+\rho_{N+k+j+1}(\nu)=0
\end{eqnarray*}

\noindent The determinants of the recurrence relation are  given
respectively by
$$\Delta_k(\nu)=[B^1_{2n+2k+2}(\nu)C_{2n+2k+1}^1(\nu)-
B_{2n+2k+2}^2(\nu)C_{2n+2k+1}^0(\nu)](n+k+\frac12)v_0^2,$$
$$\Delta_k(\nu)=[C_{2\nu+2k+2}^1(\nu)C_{2\nu+2k+1}^1(\nu)-
C_{2\nu+2k+2}^2(\nu)C_{2\nu+2k+1}^0(\nu)]v_0^2.$$
The identities in conjuction with the elementary inequality
$$\varepsilon(1-x^{\varepsilon})\leq \log x \leq
\varepsilon(x^{\varepsilon}-1)$$ allows to exlcude the annihilation
of the determinants.

\subsection{Case II. The coefficients of the expansion}

In  the preceding section we assumed  that  $v_0 \neq 0$
when the `initial power ' $\alpha=-1$.  However we mentioned 
that when the initial power $\alpha > -1$ then we should use 
the Neumann series around the Bessel operator $H_{\kappa}$.  
The preceding paragraph suggests that certain terms are excluded from the
first order perturbational terms and hence signify that the potential
is decomposed as:
$$v(x)=x^{2\nu}v_{0}+x^{2\nu}u(x) + {\tilde v}$$
where $u \in C_{0}^{\infty}(\overline{{\bf R}_{+}})$ and as
$x \rightarrow 0,$
for $u_{2j}=v_{2\nu+2j}$ then $u(x) \sim \sum_{j=1}^{\infty}u_{2j}x^{2j}$ . 
Expanding the trace we concentrate on the contribution of  the first three 
terms, abbreviating  $R_{\lambda} \equiv R_{\lambda}(H_{\kappa})$ to obtain 
$mod(\lambda^{-\infty})$: 
\begin{eqnarray*}
Tr(R_\lambda \cdot v R_\lambda \cdot v)&=&
2u_{0}Tr(R_\lambda x^{2\nu} R_\lambda \cdot \tilde{v}) +
2u_{0}Tr(R_\lambda x^{2\nu} R_\lambda x^{2\nu} \cdot u) \\
&+& u_{0}^{2}Tr(R_\lambda x^{2\nu}  R_\lambda \cdot x^{2\nu})
+ 2Tr( R_\lambda \cdot {\tilde v} R_\lambda \cdot x^{2\nu}u)\\
 &+& Tr(R_\lambda \cdot {\tilde v} R_\lambda \cdot {\tilde v}) +
Tr(R_\lambda \cdot x^{2\nu} u R_\lambda \cdot x^{2\nu} u ).
\end{eqnarray*}
However, there we should assume that $u_0=v_{2\nu} \neq 0$; otherwise
we have through the commutator identity 
$$ x^{2j} R_{\lambda}= R_{\lambda}x^{2j}+2jR_{\lambda}D_{2j-3}x^{2(j-1)}
R_{\lambda} $$ which by iteration allows us to exhaust all the $x-$
powers at the cost of $R_{\lambda}$-powers. At this point the 
scaling argument restricts the contribution to a given order only 
to that from the first term.
Therefore, the traces  are essentially comprised in the general form 
$$I(\sigma):=Tr(R_\sigma \cdot x^{2\nu} \cdot R_\sigma 
\cdot {\cal V})$$ where ${\cal V}$ stands for $x^{2\nu}u,\tilde{v}$.
The identity $$\partial_xx^{\nu+1}I_{\nu+1}(x)=x^{\nu+1}I_{\nu}(x)x$$ 
suggests further that $$\partial_xx^{2\nu+2}(I^2_{\nu+1}(x)+I^2_{\nu}(x))
=2(2\nu+1)x^{2\nu+1}I^2_{\nu}(x)$$ and an integration of the latter  
reduce the study to that of integrals of the form 
for $\xi=\frac{\sigma}{x}$: $$I(\sigma)= \frac{\sigma^2}{2}
\int_{0}^{\infty}\frac{dx}{x} x^{2\nu+2}{\cal V}(x) {\cal F}(\xi) $$
where $${\cal F}(\xi)= \xi^2\frac{1}{2\nu+1}
(\xi^{-2}B_{0}(\frac1\xi)+ B_{1}(\frac1\xi))$$
having set that  
$$ B_{j}(\xi):=(\xi^{j}K_{\nu}I_{\nu+j})(\xi)\cdot 
(\xi^{j}K_{\nu}I_{\nu+j})(\xi), $$
The Singular asymptotics lemma  suggests that if $2\nu+\alpha \notin {\bf Z}$
$v_{\alpha}\neq 0$ is a nonvanishing 
asymptotic coefficient of the ``potential ''  ${\cal V}$ then
$$ D_{2\nu +\alpha+2,0}I (0)=C_{2\nu+\alpha+2}(\nu)u_{0}v_{\alpha} + 
P_{2\nu +\alpha+2}(v)$$  while if $2\nu+\alpha \in {\bf Z}$ then 
$$ D_{2\nu+\alpha+2,1}I(0)=B_{2\nu+\alpha+2}(\nu)u_0v_{\alpha} + 
Q_{2\nu+\alpha+2}(v).$$ For these we need the Mellin tranfrom of 
${\cal F}(\xi)$ which are calculated in the next paragraph since 
$$C_{2\nu+\alpha+2}(\nu)= u_{2\nu+\alpha+2,0}({\cal F})=
\widehat{\cal F}(-2\nu-\alpha-2),$$
$$ B_{2\nu+\alpha+2}(\nu)=D_{2\nu+\alpha+2,0}{\cal F}=
Res_{s=-(2\nu+\alpha+2)}(\widehat{\cal F}) .$$

\subsection{Bessel function formulae.} 

We give in some detail the calculation of the Mellin transform of the
functions: ${\cal B}_{0},{\cal B}_{1}$.
The calculations of the Mellin transforms are executed 
using the integral representation for $j=0,1$
$$B_j(\xi)=\xi^{j}K_{\nu}I_{\nu+j}(\xi)=\int_{0}^{\infty} 
\frac{x^{j+2}}{x^2 +\xi^{2}}J_{\nu+j}(x)J_{\nu}(x)\frac{dx}{x}$$
and the classical Weber-Schaftheitlin 
integral\footnote{ These are condensed in one formula: 
 $$\int_{0}^{\infty}x^{j-s}J_{\nu+j}(x)J_{\nu}(x)
\frac{dx}{x}=  \frac{\Gamma(\frac s2+\frac12)\Gamma(-\frac s2 +\nu+j)}
{\Gamma(\frac12 s+1)\Gamma(\frac s2 +1+\nu)}$$}.
The convolution formula allows thus to obtain for:
$$ \widehat{{\cal B}_{j}}(s)= 
\frac{(\Gamma(j+\nu-\frac s2 ))^{2}}
{8\sqrt{\pi}\Gamma(\frac12 +\nu)^{2}(2\pi i)}  \cdot
\int_{\lambda-i\infty}^{\lambda+i\infty}
 F_{1}(w)F_{2}(w)F_{3}(w)F_{4}(w)dw$$
 setting 
$$ F_{1}(w):=\frac{\Gamma(\frac{w-s}{2} )\Gamma(-\frac w2 +\nu+j)}
{\Gamma(-\frac s2+\nu+j)}, \hspace{0.5truecm}
F_{2}(w):=\frac{\Gamma(-\frac w2 )\Gamma(\frac{w-s}{2}+\nu+j)}
{\Gamma(-\frac s2 +\nu +j)},$$
$$ F_{3}(w):=\frac{\Gamma(\frac{s-w+1}{2})\Gamma(\frac12 +\nu)}
{\Gamma(\frac{s-w}{2}+1 +\nu  )}, \hspace{0.5truecm}
F_{4}(w):=\frac{\Gamma(\frac{w+1}{2})\Gamma(\frac12 +\nu)}
{\Gamma(\frac w2+1+\nu)}.$$
Next we study the meromorphic properties of the function $\widehat{B_{j}}$:
the  resultant formula implies in the domain
$D_{3}=\{(u,v,w) \in (\overline{{\bf R}_{+}})^{3}/ w \leq 1, uv \geq w \}$ that:
$$\widehat {{\cal B}_{j}}(s)=\frac{(\Gamma(j+\nu-\frac12 s))^{2}}
{\sqrt{\pi}(\Gamma(\frac12 +\nu))^{2}} 
\cdot I_j(s)$$ where  $$I_j(s):=\int_{D_{3}}\frac{dudvdw}{uvw}
[u^{-s}((1+u^2)^{-\nu + \frac s2})]
[v^{-s+2\nu+j}(1+v^2)^{-\nu -j+\frac s2}]  w^{s+1}(1-w^2)^{\nu-1/2}
\cdot $$ $$ \cdot (\frac{w}{uv})\cdot
(1-\frac{w^2}{u^2v^2})^{\nu-1/2}$$ 
This is written further through a Taylor development as 
$$I_j(s)=\sum_{l=0}^{\infty}
\frac{\Gamma(\frac s2 +\frac32+l)}{l!\Gamma(\nu+\frac12 -l)
\Gamma(\frac s2+\nu+2+l)}I_{jl}(s)$$ 
if we introduce the integral 
$$I_{jl}(s):= \int\int_{uv \geq 1} \frac{dudv}{uv}
[u^{-s-2l-1}(1+u^2)^{-\nu + \frac s2})] [v^{-s+2\nu+j-2l}(1+v^2)^{-\nu -j+
\frac s2}]
[u^{2\nu}v^{-(2\nu +1)} +1].$$ This represents a holomorphic function
in the left half plane $\Re s < 0$ positive on the negative
real axis. Furthermore the substitution $u^2=\frac{1}{1-\xi} -1, 
v^2=\frac{1}{\eta\xi}-1$ leads to the familiar type of integral
\footnote{ If $s=4\nu+j$ or $s=6\nu+j-1$ this integral is computed
explicitly. We omit the result since this case falls out of our goals. } 
\begin{eqnarray*}
I_{jl}(s) &=&\frac14
\int_0^1\int_0^1\frac{d\xi d\eta}{\xi\eta} \xi^{-\frac s2-\frac12+\frac j2}
\eta^{l+\frac j2} 
(1-\xi)^{l+\nu-\frac12}
(1-\eta\xi)^{-\frac s2+\nu-l+\frac j2-1}\\
&\cdot&[1+\frac{\eta^{\nu+\frac12}\xi^{2\nu+\frac12}}
{(1-\xi)^{\nu}(1-\eta\xi)^{\nu+\frac12}} ]
\end{eqnarray*}
The following facts allow us to conclude that there exists $\alpha >-2$ such 
that one of the coefficients $B_{2\nu+\alpha+2}(\nu)$ or 
$C_{2\nu+\alpha+2}(\nu)$ is non zero. These are indeed checked easily:

\begin{enumerate}         
						
\item {\it Notice that $|I_1(s)|\leq \frac12 |I_0(s)|$.

\item  For $s \in {\bf C},\Re s < 0, l \in {\bf N}$ 
we observe that $ |I_{jl}(s)| \leq I_{j0}(\Re s)$; precisely
 $$\frac{1}{(1+M)^{2l+2\nu+1}}  \leq
\frac{I_{jl}(s)}{I_{j0}(s)} \leq 1+\frac{1}{(1+M)^{2l+1+s}} $$

\item Actually as $l \rightarrow \infty$ then
 $$\frac{\Gamma(\frac s2 +\frac32+l)}{l!\Gamma(\nu+\frac12 -l)
\Gamma(\nu+2+\frac s2+l)} = O(l^{-1-\nu}).$$ The series converges
absolutely for all $ s \in {\bf C}$.

\item The function $I(s)$ has poles at the negative of 
an odd integer.}
\end{enumerate}

\begin{appendix}

In the terminology introduced already we have the following 
lemma that has appeared in various forms in the literature, \cite{c1},
\cite{c3},\cite{c4},\cite{ch2},\cite{bs} as as well as Melrose
refers to it as the push forward lemma.

\vspace{0.5truecm}
\noindent
{\bf Theorem. (The singular asymptotics lemma)}
Let $f(y,x) \in \Gamma^{\infty}({\bf R}_+^2)$. Let $(S,S')$ be the asymptotic
character of $f$. Suppose that $f$ has compact $x-$support and 
$$|\overline{P^x_{z}}[S']f(y,x)| \leq (xy)^{\Re z +\delta}h_z(y)$$ 
for some $\delta=\delta_z >0 $ and $h_z$ satisfying
$\int_0^1h_z(\frac1t)\frac{dt}{t} < \infty$. Let 
$$ F(s)=\int_{0}^{\infty}f(\frac sx,x)\frac{dx}{x}.$$
Then $F \in \Gamma^{\infty}({\bf R}_+)$ and  $S_1+S_2$ is an asymptotic
character of $F$. The asymptotic coefficients of $F$ are given by 
\begin{eqnarray*}
D_{k,j}F(0)&=&
\sum_{r=j}^{S(k)-1}\frac{r!}{j!}u_{k,r-j}^xD_{k,r}^{y}f(y,x)|_{y=0}
+\sum_{r=j}^{S'(k)-1}\frac{r!}{j!}u_{k,r-j}^yD_{k,r}^{x}f(y,x)|_{x=0}\\
&-& \sum_{r=0}^{j-1}\frac{(j-r-1)!r!}{j!}D_{k,r}^xD_{k,j-1-r}^{y}f(y,x)|_{
x=y=0}
\end{eqnarray*}
where $u_{k,j}$ is a linear functional on $\Gamma^{\infty}({\bf R}_+)$ 
defined as follows. Let $S$ be an asymptotic character of $f$. Let
$l \geq S(k)$ and $D_z=x\partial_x-z$. Let 
$$r_k(f)(x)=f(x)-\sum_{\Re z \leq \Re k, z \neq k}\sum_{j \in {\bf Z}_+}
D_{zj}f(0)x^z\log^jx .$$ We define 
$$u_{k,j}(f)=\frac{1}{(j+l)!}\int_{0}^{\infty}x^{-k}(-\log^{j+l}x)
r_{k}(D_k^lf)(x)\frac{dx}{x} .$$ It can be readily checked that $u_{kj}(f)$
is independent of $l$ and therefore of $S$, for $S$ large enough.

\vspace{0.5truecm}
\noindent
The computations are in fact facilitated through Mellin transforms because
of the formula, \cite{c4}:
$$u_{k,j}(f)=\hat{f}_{j}(-k),\hspace{1truecm}  D_{k,j}=\hat{f}_{-j-1}(-k)$$
where $f_{l}(z_0)$ is the coefficient of the $(z-z_0)^j$ in the Laurent
expansion of the meromorphic extension of the Mellin transform:
$$\hat{f}(z)=\int_0^{\infty}f(x)x^{z}\frac{dx}{x}$$ 
(which is defined for $Re z >> 0 $ if $f$ is of bounded support) around
$z=z_0$.

\end{appendix}

\bibliographystyle{my-h-elsevier}

\begin{thebibliography}{10}







\bibitem{ahs1} Avron  J. E., Herbst I. W., Simon B, {\it Schro\"dinger
operators with magnetic fields I: }, Duke Math. Jour. ,{\bf 45}, 4, 847-883,
1978


\bibitem{ahs2} Avron  J. E., Herbst I. W., Simon B, {\it Separation of
 center of mass in homogeneous magnetic fields}, Ann. of Physics,{\bf 114},
 431-451, 1978

\bibitem{bs} Br\"uning J., Seeley R. T., {\it  Regular singular
asymptotics}, Adv. Math., {\bf 58}, (1985), (1985), 133-148.



\bibitem{c1} Callias C.J., {\it  Spectrally determined potentials}, CPDE,
 {\bf 20} , (1995), 1553-1587.


\bibitem{c2} Callias C. J., {\it The heat equation with singular 
coefficients}, CMP, {\bf 88} , (1983),  357-385.

\bibitem{c3} Callias C. J., {\it  The heat kernel as a function of the 
coefficients of the potential: I. Differential calculus of power-log 
asymptotic expansions},  CWRU, preprint. 

\bibitem{c4} Callias C. J., {\it Small time heat expansions}, MRL, {bf 2},
 (1995), 1-17.

\bibitem{c5} Callias C. J., {\it Private Communication}, (1994-1999).

\bibitem{ct} Callias C. J., Taubes C. H., {\it  Functional determinants of
Euclidean Yang-Mills theory}, CMP,  {\bf 77} , (1980), 229-250.

\bibitem{cu} Callias C. J., Uhlmann G. A., {\it  Singular asymptotics
approach to partial differential operators}, Bull. A. M. S., Res. Ann.,  
{\bf 11} , (1984),  172-176.
 

\bibitem{ch1} Cheeger J., {\it Analytic torsion and heat equation}, 
Ann. Math., {\bf 109} , (1983), 259-322.

\bibitem{ch2} Cheeger J., {\it  Spectral geometry of singular riemannian
 spaces}, JDG, {\bf 18} , (1983), 575-657.

\bibitem{cht} Cheeger J., Taylor M., {\it  On the diffraction of waves by
conical singularities}, I-II, CPAM, {\bf 25}, (1982), 275-331, 487-529.
Bull. A. M. S. {\bf 9} , (1983), no. 2, 315-318.

\bibitem{ds} Dunford N., Schwarz J., Linear Operators II: Spectral Theory,
 (1988) Wiley Interscience.



\bibitem{gm} Grosse H., Martin A., {\it Exact results  on potential models 
for the quarkonium models}, Phys. Reports, {\bf 60}, (1980), 341-402. 

\bibitem{gr} Gradshteyn I. S., Ryzhik I. M., Table of integrals, products and
series, Academic Press, Fifth Edition, 1994.


\bibitem{ll} Landau L. D., Lifshitz E. M., Quantum Mechanics, 
Pergamon Press, 1965


\bibitem{rs} Reed M., Simon B.,  Methods of Modern Mathematical Physics,
vol I., III, Academic Press, New York, 1972. 

\bibitem{t1} E. C. Titshmarsh, The Theory of Functions, Oxford, 
2nd edition, 1939.
 
\bibitem{t2} E. C. Titshmarsh,  The Theory of Fourier Integrals, 
Chelsea, 2nd edition, 1986.


\bibitem{t3} E. C. J. Titshmarsh, Eigenfunction Expansions
associated with second order differential operators, Oxford, 1962. 


\bibitem{w} G. N. Watson , A treatise on the theory of 
Bessel functions,  Cambridge, 1941.

\bibitem{wi} E. Witten, {\it Global gravitational anomalies}, 
Comm. Math. Phys. {\bf 100}, (1985), 197-229.


\end{thebibliography}

\end{document}